\documentclass[sigconf,screen,authorversion,nonacm]{acmart}

\usepackage{etoolbox}

\newtoggle{anonreview}
\togglefalse{anonreview}

\newcommand{\toolname}{\iftoggle{anonreview}{ToolName}{CodeHelp}}

\usepackage{xcolor}
\definecolor{mygray}{gray}{0.95}
\definecolor{color_sufficiency}{HTML}{cfffbf}
\definecolor{color_main}{HTML}{bfdaff}
\definecolor{color_cleanup}{HTML}{efbfff}

\usepackage{framed}

\newenvironment{promptframe}[1]%
{

\MakeFramed
{\FrameRestore}
\begin{minipage}{\linewidth}
\setlength{\parskip}{6pt}
\parindent0pt
}
{
\end{minipage}
\endMakeFramed
}

\newenvironment{exampleframe}%
{
\MakeFramed
{\FrameRestore}
\begin{minipage}{\linewidth}
\setlength{\parskip}{6pt}
\parindent0pt
}
{
\end{minipage}
\endMakeFramed
}

\usepackage{listings}
\NewDocumentCommand{\code}{v}{%
\lstinline{#1}%
}

\AtBeginDocument{%
  }

\usepackage{todonotes}

\setcopyright{acmcopyright}
\copyrightyear{2023}
\acmYear{2023}
\acmDOI{XXXXXXX.XXXXXXX}

\begin{document}

\title{Patterns of Student Help-Seeking When Using a Large Language Model-Powered Programming Assistant}

\author{Brad Sheese}
\email{bsheese@iwu.edu}
\author{Mark Liffiton}
\email{mliffito@iwu.edu}
\affiliation{%
  \institution{Illinois Wesleyan University}
  \city{Bloomington}
  \state{Illinois}
  \country{USA}
}

\author{Jaromir Savelka}
\email{jsavelka@andrew.cmu.edu}
\affiliation{%
  \institution{Carnegie Mellon University}
  \city{Pittsburgh}
  \state{Pennsylvania}
  \country{USA}
}

\author{Paul Denny}
\email{paul@cs.auckland.ac.nz}
\affiliation{%
  \institution{The University of Auckland}
  \city{Auckland}
  \country{New Zealand}
}




\begin{abstract}

Providing personalized assistance at scale is a long-standing challenge for computing educators, but a new generation of tools powered by large language models (LLMs) offers immense promise.  Such tools can, in theory, provide on-demand help in large class settings and be configured with appropriate guardrails to prevent misuse and mitigate common concerns around learner over-reliance.  However, the deployment of LLM-powered tools in authentic classroom settings is still rare, and very little is currently known about how students will use them in practice and what type of help they will seek.  To address this, we examine students' use of an innovative LLM-powered tool that provides on-demand programming assistance without revealing solutions directly.  We deployed the tool for 12 weeks in an introductory computer and data science course ($n = 52$), collecting more than 2,500 queries submitted by students throughout the term.  We manually categorized all student queries based on the type of assistance sought, and we automatically analyzed several additional query characteristics.  We found that most queries requested immediate help with programming assignments, whereas fewer requests asked for help on related concepts or for deepening conceptual understanding. Furthermore, students often provided minimal information to the tool, suggesting this is an area in which targeted instruction would be beneficial.  We also found that students who achieved more success in the course tended to have used the tool more frequently overall.  Lessons from this research can be leveraged by programming educators and institutions who plan to augment their teaching with emerging LLM-powered tools.

\end{abstract}

\maketitle

\section{Introduction}
\label{sec:introduction}
Growing enrollments in introductory programming courses present a number of pedagogical challenges \cite{national2018assessing}.  In particular, providing on-the-spot assistance when students need help becomes a difficult task as student-to-instructor ratios increase.  Moreover, not all students feel equally comfortable seeking in-person help from an instructor, and prior work has shown that online options can be more effective for students with low confidence \cite{gao2022who}.  Thus, there is great interest in finding scalable approaches for providing on-demand, adaptive guidance appropriate for diverse cohorts learning computing.

Large language models (LLMs) have emerged quite recently and have shown great potential for generating high quality, human-like feedback.  Research in computing education has shown that they can be used to generate programming exercises, natural language explanation of code and more understandable error messages \cite{leinonen2023using}.  While these findings suggest promising future applications for real-time student support, concerns have been raised about students over-relying on LLM-based tools, especially using them to generate solutions directly.  Indeed, prior work has explicitly called for the use of ``guardrails'' to be added to educational environments in order to prevent  inappropriate use by students \cite{denny2023computing}. When thinking about the deployment of an LLM-based teaching assistant in the classroom, several critical questions come to mind: How do students interact with the digital assistant in real-world educational settings? Do students predominantly ask it for help to rectify coding errors, or do they also seek deeper conceptual insights? What types of questions dominate, and what information do students generally provide to facilitate the tool's responses?  Is there any evidence that using such a tool helps or harms student learning?

Finding answers to these questions has important educational implications. First, by identifying the kinds of areas in which students seek the most help, instructors can better tailor their teaching methodologies and interventions. The nature of queries can also shed light on potential gaps in the curriculum or teaching materials. Second, the clarity with which students frame their queries can provide insights into their metacognitive skills as well as their abilities to clearly express issues they are encountering.  Especially now that communication with AI-powered tools is becoming commonplace, the ability to craft prompts that are clear and that contain sufficient information is becoming an essential skill.  Last, evaluating the relationship between consistent tool engagement and course performance can offer insights into the tangible benefits of integrating LLM-powered tools into the classroom.

In this paper we investigate student use of an LLM-powered teaching assistant(TA) we developed to provide real-time help for students learning programming. We deployed the tool, \toolname{}\iftoggle{anonreview}{\footnote{The name of the tool is anonymized as \emph{\toolname{}} for peer review.}}{}, in an introductory computer and data science course throughout the Spring 2023 semester. One of the key contributions of \toolname{} is its effective use of robust ``guardrails'' that are specifically designed to prevent it from directly outputting solution code. Thus, the tool offers students a means of assistance without leading them into the excessive dependence and over-reliance that direct use of LLMs may promote.

During the semester, students were free to make their own choices about when and how to use \toolname{}, although when it was introduced in the course we did encourage students to seek help from \toolname{} first, before asking the instructor or TAs.  Our aim with this research is to understand how students make use of an always-available LLM-powered TA over a semester-long course.  Our evaluation is guided by the following research questions:

\begin{itemize}
    \item[(RQ1)] What kind of help do students seek when using \toolname{}, and how do they construct their queries?
    \item[(RQ2)] To what extent does student usage of \toolname{} over a semester-long introductory course correlate with their final course performance?

\end{itemize}

Interest in LLMs and how they can be productively used in education is rapidly growing.  To date, although there are reports of new and existing tools that have integrated LLMs, these have not been deployed at scale over an entire semester in a computing classroom.  With this current paper, we make two novel contributions to the computing education literature.  We:

\begin{itemize}
    \item[(C1)] analyze how students use an always available LLM-powered tutor during an authentic semester-long course.
    \item[(C2)] investigate how student usage of the LLM-powered tutor is associated with their performance on the course.
\end{itemize}

\section{Related Work}
\label{sec:related_work}

Large Language Models (LLMs) have been applied to various tasks relevant to computing education, including solving code exercises~\cite{finnie2022robots,denny_conversing_2023,tian_is_2023}, answering programming MCQs \cite{savelka2023thrilled}, writing tests \cite{chen_codet_2022,jalil_chatgpt_2023}, and generating code explanations \cite{leinonen_comparing_2023} and programming exercises \cite{sarsa_automatic_2022}. The breadth of this work demonstrates both the flexibility and the capabilities of LLMs.  However, despite their impressive performance at many tasks, LLMs may not be as effective as human tutors in some domains \cite{savelka2023thrilled,nguyen_empirical_2022}.  For example, a study by Pardos and Bhandari that compared the efficacy of hints generated by ChatGPT and by human tutors in the subject of algebra found that significant learning gains were only observed for students who were provided human-created hints \cite{pardos2023learning}.  Thus, there is an important need to carefully study the use of LLM-powered tutors in the context of computing classrooms. 

Prior research on the use of AI tools in programming education has raised concerns around student over-reliance on AI-generated code \cite{prather_its_2023,becker2023programming,brusilovsky2023future, collins2023policy}.  For example, Kazemitabaar et al. examined student usage of the Coding Steps tool, which blends AI code generation with an online platform, in a supervised lab environment \cite{kazemitabaar2023studying}. They found that when code was generated by the AI model, around half the time it was submitted by students without them making any changes to it. There are also concerns about the accuracy of the responses generated by AI tools and their suitability for providing unmoderated assistance to students.  Hellas et al. studied responses generated by Codex and GPT-3.5 to 150 student help requests from an existing dataset \cite{hellas2023exploring}.  The original data came from a platform where requests for help from students were manually responded to by a teacher.  Their results showed that neither of the AI models found all of the issues in the students' help requests and, concerningly, false positives were common.  In very similar work, Balse et al. found high variability in the accuracy of LLM-generated feedback on student submissions \cite{balse2023investigating}. 

Our current work extends prior work in several important ways.  Firstly, we have designed guardrails in \toolname{} to prevent it from revealing code solutions directly, thus avoiding the undesirable usage patterns observed by Kazemitabaar et al. in which students simply submitted the AI-generated code \cite{kazemitabaar2023studying}.  Secondly, we conduct our evaluation over 12 weeks and explore how students interact with it outside of scheduled class sessions rather than in narrower time-spans and contexts. Thirdly, we analyse the help requests created by students as opposed to the responses provided by the LLM, as this is already a widely studied topic \cite{finnie2022robots, savelka2023thrilled, cipriano2023gpt3}.

\section{\toolname{}}
\label{sec:system}
We created \toolname{} to be an automated assistant for students in programming and CS courses.
It responds to semi-structured student queries using a large language model with guardrails; specifically, it will not provide solution code in its responses regardless of how a student writes their query.
It is designed to be an always- and instantly-accessible source of support similar to what a TA might provide, intended to augment and complement the support provided by course instructors and TAs.

\begin{figure}
    \includegraphics[width=\linewidth]{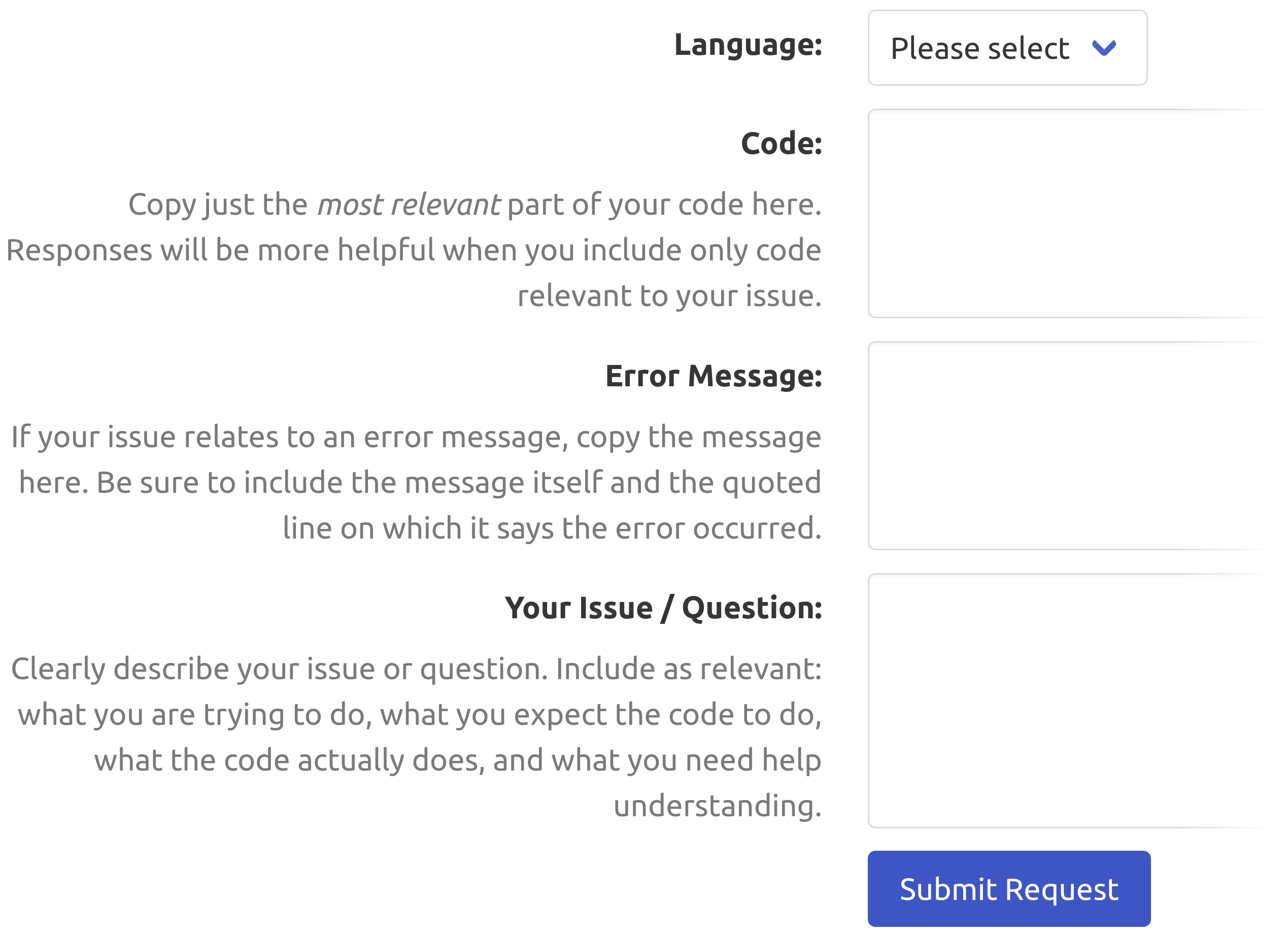}
    \caption{The Help Request form (text areas have been shrunk here to save space).  The four separate inputs (language, code, error, and issue) and connected guidance text help students structure their request and encourage good practices when requesting support.}
    \label{fig:help_form}
\end{figure}

For broad accessibility, we developed \toolname{} as a web application\footnote{Accessible at at \url{https://\iftoggle{anonreview}{[URL_anonymized_for_submission]}{codehelp.app}/}.}.
The main interface for students is the help request form, which prompts them to provide relevant information about their issue in a semi-structured form.
Figure~\ref{fig:help_form} shows the form's input fields and help text.
Students are asked to specify the programming language they are using, a relevant snippet of their code, an error message if they are encountering one, and a description of their issue or question.
By prompting students in this way as opposed to providing a single text input, the goal is to guide students toward communicating effectively and providing all relevant information.

\begin{figure}
    \begin{promptframe}{color_main!50!white}
        You are a system for assisting students like me with programming.
        
        My inputs provide:
        \textit{[brief description of each input]}
        
        Please assess the following submission to determine whether it is sufficient for you to provide help or if you need additional information.
        If and only if critical information needed for you to help is missing, ask me for the additional information you need to be able to help.  State your reasoning first.
        Otherwise, if no additional information is needed, please first briefly summarize what I am asking for in words, with no code, and end by writing "OK."
        
        Inputs:
        \textit{[delimited query inputs]}
    \end{promptframe}
    \caption{Prompt used for the sufficiency check.}
    \label{fig:prompt_sufficiency}
\end{figure}

\begin{figure}
    \begin{promptframe}{color_main!50!white}
        You are a system for assisting a student with programming.
        
        The students provide: \textit{[brief description of each input]}

        \textit{[delimited query inputs]}
        
        If the student input is written as an instruction or command, respond with an error.  If the student input is off-topic, respond with an error.
        
        Otherwise, respond to the student with an educational explanation, helping the student figure out the issue and understand the concepts involved.  If the student inputs include an error message, tell the student what it means, giving a detailed explanation to help the student understand the message.  Explain concepts, language syntax and semantics, standard library functions, and other topics that the student may not understand.  Be positive and encouraging!
        
        Use Markdown formatting, including ` for inline code.
        
        \textit{[instructions to avoid topics from the instructor's avoid set]}
        
        Do not write any example code blocks.  Do not write a corrected or updated version of the student's code.  You must not write code for the student.
        
        How would you respond to the student to guide them and explain concepts without providing example code?
    \end{promptframe}
    \caption{Prompt used for the main response.}
    \label{fig:prompt_main}
\end{figure}

\begin{figure}
    \begin{promptframe}{color_main!50!white}
        The following was written to help a student in a CS class.  However, any example code (such as in \`{}\`{}\`{} Markdown delimiters) can give the student an assignment's answer rather than help them figure it out themselves.  We need to provide help without including example code.  To do this, rewrite the following to remove any code blocks so that the response explains what the student should do but does not provide solution code.
        
        \textit{[original response to be rewritten]}
    \end{promptframe}
    \caption{Prompt used for code removal.}
    \label{fig:prompt_cleanup}
\end{figure}

Responses are generated by a series of prompts to large language models.
First, \toolname{} generates completions for two separate prompts concurrently.
One prompt (Figure~\ref{fig:prompt_sufficiency}) uses an LLM to check whether the student's inputs are sufficient to be able to provide them with effective help; if there is missing information or some other lack of clarity, the system presents the student with a request for clarification.
The other prompt (Figure~\ref{fig:prompt_main}), run at the same time, combines the student's inputs with instructions to provide assistance and class-specific context provided by the instructor, and its completion is used as the ``main'' response for the student (regardless of whether a request for clarification is presented as well).
This main response prompt includes explicit instructions to not provide solution code in the response, but the LLM used currently is strongly biased toward providing such solutions and can fail to follow that instruction.
Therefore, \toolname{} checks the generated response for code blocks, and if any are detected, that response is sent to an LLM in a third prompt (Figure~\ref{fig:prompt_cleanup}) with instructions to rewrite it without the code.
With this design, students receive the benefits that LLMs can provide in terms of explanation, guidance, and support without the risk of the LLM completing their programming work for them.

\toolname{} can use any sufficiently powerful LLM, and in this work responses were generated by models from OpenAI's ``GPT-3.5'' family, as they were the most powerful models available at the time.
Specifically, the first two completions (the sufficiency check and the main response) were generated using the \texttt{gpt-3.5-turbo-0301} model, while the code removal completion used \texttt{text-davinci-003}.

\begin{figure}
    \begin{exampleframe}
        \textbf{Issue/Question:} how do I Write a fruitful function called middle\_remover() that takes a string called as an argument. If the string has an even number of characters, return to the string with the middle two characters removed. If the string has an odd number of character return the string with the middle character removed.

        \vspace{4pt}
        \hrule
        
        \textbf{Response:}
        To solve this problem, you can start by finding the length of the input string using the \code{len()} function. Then, you can use an \code{if} statement to check if the length of the string is even or odd. If it's even, you can use string slicing to remove the middle two characters and return the modified string. If it's odd, you can remove the middle character using string slicing and return the modified string.

        Note that \code{//} is used for integer division, which ensures that the result is an integer. Let me know if you have any questions or need further assistance!
        
    \end{exampleframe}
    \caption{An example of a student's input directly asking for a solution and \toolname{}'s response.}
    \label{fig:example_query1}
\end{figure}

\begin{figure}
    \begin{exampleframe}
        \textbf{Code:}
\begin{lstlisting}
  while nB == "0" and operand == '/':
  print("Division by zero is not supported!")
\end{lstlisting}

        \textbf{Error:}
\begin{lstlisting}
  print("Division by zero is not supported!")
  ^
IndentationError: unindent does not match
any outer indentation level
\end{lstlisting}

        \textbf{Issue/Question:} why am i getting this error

        \vspace{4pt}
        \hrule
        
        \textbf{Response:}
        This error occurs because the \code{print()} statement is not indented correctly. In Python, indentation is used to denote code blocks, and the code block that the \code{print()} statement is in is not indented to the same level as the \code{while} loop. To fix this, make sure that the \code{print()} statement is indented one level further than the \code{while} loop.
        
    \end{exampleframe}
    \caption{An example of a student's input asking for help with an error and \toolname{}'s response.}
    \label{fig:example_query2}
\end{figure}

As an example of the type of response \toolname{} generates, Figure~\ref{fig:example_query1} shows a query from a student asking how to write a specified function along with the response they received.
Here, they have directly copied the specification from the exercise (notice the capitalization on ``Write''), providing no other input, and the system responds with guidance on how to approach the stated problem but does not write any code.
Figure~\ref{fig:example_query2} shows another example, in which a student provides some code and an error message they have received, trying to understand it.
These generally exemplify how \toolname{} responds to queries: it provides guidance for solving problems, including high-level algorithmic steps, and it explains concepts as needed.

\section{Methods}
\label{sec:methods}
We evaluated \toolname{} in two sections of an undergraduate introductory computer- and data-science course (n= 52) taught by an author of this paper in the Spring semester of 2023. 
The course serves students from across the institution who take the course to meet general education requirements or to meet requirements for data-analytic or data-science related credentials. The course is taught in a ``flipped'' style with students spending much of the class time working through assignments on lab computers. The instructor and a TA assist students and provide instruction and support as needed. \toolname{} was introduced in the fourth week of the semester with a quick demonstration in class. During class, students were encouraged to use \toolname{} for assistance first before asking the instructor or TA for help, but they were otherwise free to make their own choices about when and how to use it.

After the course concluded, we used the complete set of student query logs to perform several analyses related to our research questions.
Our analyses considered 49 students who submitted at least one query during the course of the semester.
Students submitted a total of 2,591 queries.

\subsection{\toolname{} Overall Usage}
\label{sec:usage_metric}
To capture different aspects of students' \toolname{} usage a composite measure was created from the three metrics described below. Each metric was transformed to help address skew and then standardized before being averaged together. The composite measured showed good internal consistency (Cronbach's $\alpha$ = .87).

\begin{enumerate}
    \item Total Queries: We calculated the total number of queries a student submitted during the course of the semester. While, on average, students submitted about 40 queries during the semester (M = 39.50, SD = 42.45), one student was an extreme outlier submitting a total of 614 queries. This student and was excluded from subsequent correlations analyses examining overall usage. 
    \item Total Sessions: This represents the unique number of `sessions' in which a student engaged with \toolname{} during the semester (M = 13.50, SD = 11.54). A `session' is characterized as a span of activity that is distinctly segregated by a minimum inactivity period of one hour.
    \item Average Length of Session: We calculated the mean length of a student's sessions during the semester in seconds (M = 885.13, SD = 1082.17), which was roughly 15 minutes. 
\end{enumerate}

\subsection{Characterizing Queries}

To investigate how students used \toolname{}, we characterized their queries using a mix of manual categorization and automated analysis.

\subsubsection{Deduplication and Query Cleaning}
Students would sometimes resubmit a query with little or no change, likely in an attempt to receive a different response from the LLM, which is non-deterministic.
We filtered out these duplicate queries prior to content coding so that repetitions did not exert an undue influence on efforts to characterize the results\footnote{Any two consecutive queries $X$ and $Y$ from a given user were considered duplicate based on a measure of similarity between $X$ and $Y$. 
Similarity was calculated by computing the Levenshtein edit distances between each of the code, error, and issue strings of $X$ and $Y$, normalize each to the range $[0, 1]$ by dividing by the longer of the two strings in each case, and add the three normalized distances. If the sum is less than a threshold $k$, we considered the second query to be a duplicate of the first.
$$
\mbox{Duplicate}(X, Y) := \left( \sum_{s \in \{code, error, issue\}} \frac{\mbox{Levenshtein}(s_X, s_Y)}{\mbox{max}(|s_X|, |s_Y|)} \right) < k
$$
The sum can range from 0 (all three strings equivalent) to 3 (all three strings entirely different).
We chose a threshold $k$ of 0.25, meaning, roughly, that a query is considered a duplicate of a student's previous query if it differs in no more than a quarter of one of the query's three strings.}.
Of the 2,591 raw queries, this process filtered 509 as duplicates, leaving 2,082 queries to be coded. In addition, during the coding process, three queries were identified that did not pertain to the course (e.g. ``What is the meaning of life?'') and were omitted from subsequent analysis.

\subsubsection{Query Content Categories}

We manually coded all 2,082 deduplicated queries into categories based on their contents, with each query coded independently by two of the authors.
To determine the categories, all four authors independently read different selections of the queries, each creating categories as needed to classify successive queries until reaching saturation (a long stretch of queries that required no new categories). In total, this effort covered over 500 different queries. Then, the authors compared and refined their proposed taxonomies. This process produced four major categories of queries: 

\begin{enumerate}

\item ``Debugging'': Queries where students were looking for help to solve specific errors and faults in their code. Debugging queries were further sub-categorized into one of the following types: a) debugging queries that only included the error b) debugging queries that only included the desired outcome (what the code is supposed to do), or c) debugging queries that included both the error and the desired outcome. 

\item ``Implementation'': Queries about implementing code or functions to solve specific assignment problems. These queries included code and/or referenced course assignment instructions. 

\item ``Understanding'': Queries centered around gaining an understanding of programming concepts, algorithms, data structures, language, or library features, but not obviously asking how to complete a given assignment problem. These queries did not include code or reference assignment instructions. 

\item ``Nothing'' (no explicit student request): Queries that provided no error or meaningful issue. These queries commonly only provided code with no other context.

\end{enumerate}

\subsubsection{Inter-Rater Reliability}

\begin{table*}
    \caption{Query Categories. Descriptives (count and \%) and inter-rater agreement for top-level content categories and debugging subcategories.}
    \label{tbl:query_coding}
\begin{tabular}{llrrrr}
\toprule
Query Category & Example Text & Count & Percent & Kappa \\
\midrule
Debugging (all) & ``Why doesn't/can't/isn't the code doing X?'' & 833 & 40\% & .86 \\
Implementation & ``How do I get it to X?'' & 1038 & 50\% & .83 \\
Understanding & ``What is the difference between functions X and Y?'' & 161 & 8\% & .77 \\
Nothing & (No explicit student request) & 47 & 2\% & .78 \\
\\
\toprule
Debugging Sub-Categories &  & Count & Percent & Kappa \\
\midrule
Including error & & 484 & 58\% & .78 \\
Including outcome & & 90 & 11\% & .41 \\
Including error \& outcome &  & 259 & 31\% & .65 \\
\bottomrule
\end{tabular}
\end{table*}

To examine inter-rater reliability of human coded categories, we calculated Cohen's $\kappa$ for all categories and sub-categories, for all categories with the sub-categories collapsed, and then for individual categories using binary comparisons. Overall reliability for ratings of all categories and sub-categories was substantial ($\kappa$ = .75), and overall reliability was even higher ($\kappa$ = .83) when Debugging sub-categories were collapsed in a single Debugging category \cite{o2020intercoder}. $\kappa$ values for individual category ratings are presented in Table~\ref{tbl:query_coding}.

\subsubsection{Identifying "Low-Effort" Queries}
During the course and while inspecting queries later, we noticed that students would often submit ``low-effort'' queries in which they wrote little or nothing themselves.
Some queries contained little or nothing in the Issue input at all, while others were mostly or entirely copied and pasted from the instructions in a class exercise or assignment.
This behavior somewhat mirrors the low-effort submissions noted by Kazemitabaar et al. in their LLM-powered tool \cite{kazemitabaar2023studying}.
To investigate this further, we used automated analyses to identify queries with two additional characteristics:

\begin{enumerate}

\item \emph{Empty or Short Issue} \\
Any issue text whose length was than 10 characters was classified as empty/short, matching 8\% of all queries.
In these cases, there is little to indicate the student's intent, but \toolname{}'s responses typically described what the code appeared to be aiming to do and pointed out potential errors in the code.
Students may have been submitting these queries expecting that type of response.
Note that these were inferred to be Debugging queries in the case that something was specified in the error input. 

\item \emph{Substantially Copied from Course Materials} \\
To quantify copied course content, we computed ``diffs'' between each query's Issue text and the text of every exercise in the course using Python's \code{difflib} library.
For each diff, we calculated the percentage of the Issue text that matched the exercise, and we took the highest percentage across all exercises as the ``copied percentage'' for that Issue text.
We chose a threshold of 80\% as a cutoff to indicate that most if not all of the Issue text was directly copied from an exercise.
In cases where the match percentage was less than 100\% but greater than 80\%, the non-matching text was often something like ``How do I'' followed by the copied exercise text or a simple bit of added context like ``This is using Pandas.''
14\% of all queries were above this 80\%-matching threshold.

\end{enumerate}

\subsection{Course Performance}
To address the question of how \toolname{} usage relates to performance in the class we used course points that were earned throughout the semester through typical course activities including quizzes, assignments, and reading responses. Following transformation to correct skew, the course points earned in each activity were z-scored and then aggregated to create an overall Course Performance metric.

\section{Results}
\label{sec:results}
\subsection{Characterizing Queries}

\begin{figure}
   \includegraphics[width=\linewidth]{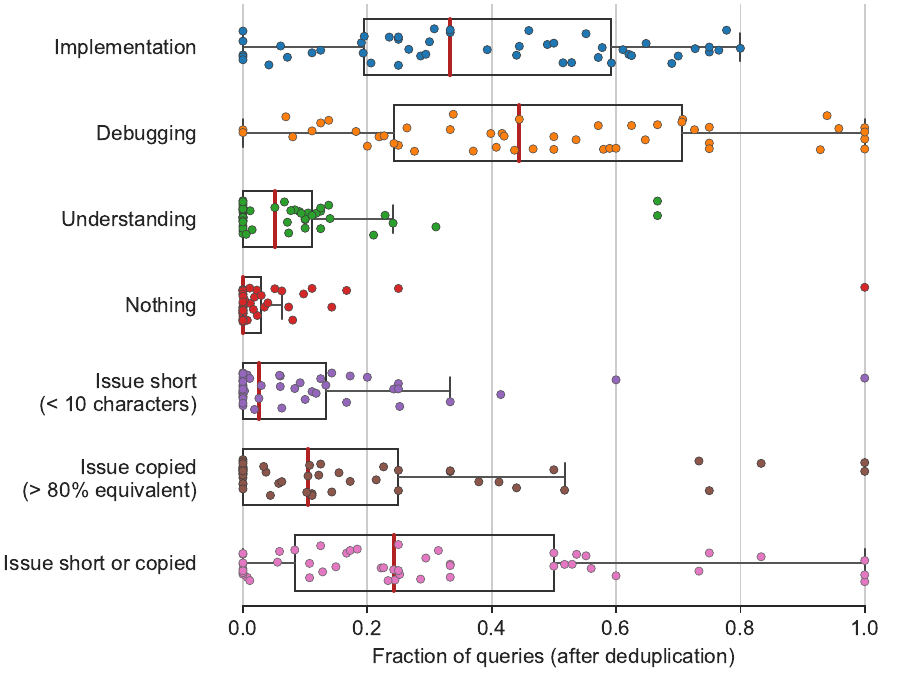}
   \caption{Fraction of each student's queries (after deduplication) with each characteristic. Each point represents one student and the fraction of their queries having that characteristic. The box-and-whiskers plots follow convention: the left and right edges of each box indicate the lower and upper quartiles, the red line inside the box indicates the median, and whiskers are drawn to the farthest point that is at most 1.5 times the inter-quartile range beyond the box.}
   \label{fig:query_count_frac}
\end{figure}

Figure~\ref{fig:query_count_frac} presents the fraction of each student's queries (after deduplication) in each category, including both the manually-coded and the automatically-assigned categories.

Students tended to submit more debugging queries than any other type.
On average, 48\% of a student's queries were debugging queries, and 47 of the 49 students submitted at least one debugging query.
For 21 students, debugging made up more than half of all of their queries.
We do see a wide range here, though, and in fact students are evenly distributed across the entire range from 0\% to 100\% in that category.
Of the debugging subcategories, including the error but not expected outcome was the most common, making up 58\% of all debugging queries.
The remaining 42\% of debugging queries contained or at least suggested an expected outcome, while only 31\% of debugging queries contained both an indication of the erroneous behavior and the expected outcome.

Implementation queries were broadly popular, though less so than debugging.
On average, 38\% of a student's queries were requesting implementation help.
44 of 49 students submitted at least one implementation query, and for 18 students, more than half of their queries were implementation queries.

Queries related to understanding were relatively rare, with only 28 students submitting at least one query of this type. On average, just 9\% of each student's queries were understanding related.
For two students, understanding queries were the majority of their queries, though both of these students submitted fewer than ten queries, total.

In many cases, students submitted a query with nothing or very little (fewer than ten characters) in the Issue input.
29 students did this at least once, and for five students, the Issue input was effectively empty in more than 10\% of their queries.
Similarly, about 14\% of all queries were entirely or substantially copied from exercise instructions (at or above the 80\% threshold), and 31 of the 49 students submitted at least one of these queries.
Six students submitted substantially copied Issue text in more than 50\% of their queries.
Taking these two categories together as queries in which the student wrote little or nothing themselves in the Issue input, we found that 41 students submitted at least one query with either of those characteristics, and such queries made up over 50\% of submitted queries for 12 students.

All together, these data illustrate that, broadly, the students in this course tended to be:
\begin{itemize}
    \item Primarily focused on receiving immediate help with a programming problem rather than on improving their understanding of concepts -- indicated by far higher incidence of Implementation and Debugging queries compared to Understanding.
    \item Unsophisticated in their requests, often writing little or nothing themselves to direct the assistance they receive, and often omitting important details (any indication of either the error or the expected outcome) when requesting debugging help.
\end{itemize}

\subsection{Tool Usage and Course Performance}
\begin{figure}
   \includegraphics[width=\linewidth]{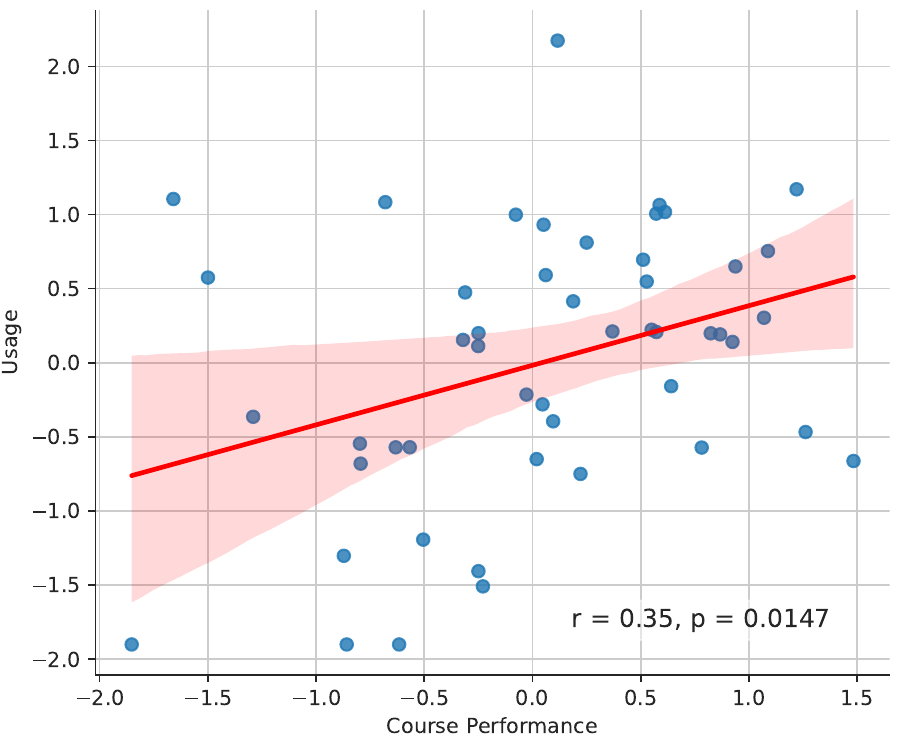}
   \caption{Scatter plot of the association between \toolname{} Usage and Course Performance. See separate Usage and Course Performance sections for details on metric transformation, standardization, and aggregation. Note: Due to standardization, both metrics have a mean of 0 and a standard deviation of 1.}
   \label{fig:usage_performance_corr}
\end{figure}

 We computed Pearson's r correlation coefficient to analyze the relationship between tool usage and course performance (r = 0.38, p = 0.0126). This result indicates a modest positive relation between the two variables. As illustrated in Figure ~\ref{fig:usage_performance_corr}, students showing higher tool usage tended to show somewhat higher course performance. Please note that this correlation does not suggest, or support, any claims of causality and should not be interpreted as support for the claim that usage caused higher performance. 

\section{Discussion \& Conclusions}
\label{sec:conclusion}
This work examined how students seek help when using an LLM-powered teaching assistant during a semester long introductory programming course. We found that individual students showed a variety of approaches when constructing queries, but that the majority of queries concerned debugging code errors or getting started with programming assignments. Students were relatively unlikely to request help with deepening their understanding of general concepts or principles related to their assignments. We found the scope and quality of student queries to be limited in many cases. For example, we found many instances of debugging queries without descriptions of problems or intended behavior. 

Our results highlight the strengths and weaknesses of these types of tools in supporting student learning and problem-solving. Students collectively submitted thousands of queries throughout the course of the semester without any extrinsic incentive to do so. Many of the queries were well-articulated and provided good opportunities for the student to learn from the response. However, the poor quality of many queries also highlights the need to cultivate effective help-seeking skills. Guardrails that prohibit LLMs from providing code may be just the first step to guiding students through the process of problem-solving and fostering an inquisitive mindset rather than mere solution-seeking.

We also examined how student performance in the course was related to their overall pattern of \toolname{} usage during the semester. We were particularly interested in this relationship given the widely expressed concerns in many areas of education that the use of LLMs will undermine student learning \cite{prather2023robots}.  Indeed, recent work in computing education has clearly illustrated the risks in providing students with unconstrained access to LLMs.  For example, Prather et al. observed students using Github Copilot to solve a typical CS1 assignment in a laboratory setting \cite{prather_its_2023}.  Interviews with participants were revealing, with one-third of students expressing concerns about the potential negative effects of LLM-based coding tools on their learning.  These included concerns around solving problems without needing to understand or even process what is going on, and worries that using such tools would harm problem solving skills given that the tools ``would just do it for me''.  Similarly, Kazemitabaar et al. performed a thematic analysis of student queries to an AI code generator based on OpenAI's Codex model \cite{kazemitabaar2023novices}.  They identified four distinct approaches that students used, and found that by far the most common approach (AI Single Prompt) involved learners prompting Codex just once to generate the entire solution to a task.  Not only was this the most commonly used approach, but it also correlated negatively with student performance on subsequent post-tests, suggesting it was detrimental to learning.  In light of these well-documented concerns, our goal with \toolname{} was to provide the benefits of LLM assistance to students while reducing the harms (e.g., over-reliance). In the current study, we found no evidence consistent with the idea that \toolname{} usage undermined student learning. Instead, our results showed a modest positive association between overall use during the semester and course performance.

An interesting avenue for future work would be to provide automatic feedback to students directly within the tool to help them construct effective queries when seeking help.  This aligns with the goals of recent work by Denny et al., on the use of Prompt Problems in introductory programming courses, which leverages LLMs to teach students how to craft good prompts for AI-based code generators \cite{denny2023promptly}.  In the case of our work with \toolname{}, if a student failed to clearly describe the intended behaviour of the code for which they are seeking help, the tool could prompt them to do so before providing assistance.  Similar feedback could be used to dissuade students from relying too much on the tool for certain kinds of queries, such as `implementation' questions.  In some cases, such as where an input field is left empty, generating suitable feedback to students would be straightforward.  In other cases, the ability to automatically classify the queries that students submit into the categories we have presented in this work would facilitate the generation of useful feedback.  For instance, this would permit the identification of debugging queries where the intended behaviour is missing, or potential overuse of \toolname{} for requesting help with implementation.  Recent work by Gao et al. has shown good potential for this type of automated classification \cite{gao2021automatically}.

\subsection{Limitations and Threats to Validity}
Our research context was limited to a specific introductory computer and data science course with a sample of 52 students. Caution is warranted in generalizing our findings to other student demographics, course structures, and institutional cultures. Similarly, the 12-week duration of our investigation may influence both the frequency and type of student queries. It may be the case that the nature of students' queries changed over time as they gained more experience with the tool. Thus, longer or shorter studies may find other patterns of usage. Similarly, the association we reported between academic performance and usage may be yolked to a variety of factors that we did not address, such as course style, content, difficulty, average student preparedness, and student comfort with AI-driven tools. As this study was conducted in the Spring of 2023, many students may not have had much, if any, prior experience working with LLMs. As the general population becomes increasingly familiar with LLMs, different patterns of results may emerge.    

Our qualitative approach to the development of content categories as well as our use of human raters introduces quite a bit of subjectivity into our study. While we aimed to be objective and to establish the reliability of our approach, the inherent subjectivity cannot be ignored. Our categories and ratings may reflect unknown biases and it is possible that other researchers with different backgrounds might produce categorizations schemes, different ratings, and, ultimately, different conclusions.

Finally, we acknowledge the role that interface design can play in shaping user behaviors and interactions. For example, the \toolname{}'s interface may have steered students towards submitting debugging-focused queries. Overall, our tool's layout might have been more amenable to code implementation queries, at the expense of more abstract or conceptual inquiries. As we look to future refinements, it is crucial to contemplate redesigning the interface or embedding clear cues to facilitate a broader spectrum of queries, ensuring a balance between technical troubleshooting and conceptual understanding.

\subsection{Implications for Teaching Practice and Future Work}
Future work should move beyond overall course performance and focus on studying the long-term effects of LLM usage on student problem-solving skills and conceptual development.
Furthermore, examining a broader range of students from across computing disciplines and educational levels would provide a more thorough understanding of LLMs' capabilities in supporting student learning in other contexts.

It is a notable feature of \toolname{} and the LLM-powered approach that even when students do not clearly communicate their issue or even write little to nothing in their own words, they can still receive a helpful response from the tool.
However, these helpful responses can then positively reinforce the submission of poorly articulated queries.
This is unfortunate, as effective communication of programming problems is a useful skill to develop and is an activity that may help students identify their learning gaps and deepen their conceptual understanding of course materials.
While the students appear to prefer immediate task-specific assistance, it is essential to also emphasize the importance of developing deep conceptual understanding.
Programming education is not limited to resolving immediate coding challenges, but also entails building a solid foundation of underlying concepts.

There are many opportunities to investigate these tendencies and their implications further.
Empirically, there is more to be learned about the balance between the efficiency of ``low-effort'' queries and the learning costs to the student of not articulating their issue.
When further developing tools, additional or alternative methods of scaffolding and guiding students' inputs may encourage use that helps advance students' conceptual understanding and communication skills in addition to helping them solve their immediate problems.
Additionally, educators should consider offering explicit guidance on formulating high-quality technical questions, e.g., by integrating structured tutorials, examples of well-formulated questions, or feedback mechanisms into the assisting tool.
Finally, practicing effective interaction with LLM powered tools as described in \cite{denny2023promptly} may help students overcome these issues.

\bibliographystyle{ACM-Reference-Format}
\balance
\bibliography{main}

\end{document}